\documentclass[12pt]{article}		
\usepackage[super,sort,comma]{natbib}

\usepackage{fancyhdr}		

\newcommand{\captionv}[3]{\begin{center}\parbox{#1cm}{\caption[#2]{{\sf #3}}}
        \end{center}}

\usepackage{graphicx}
\usepackage{amssymb}
\usepackage{amsmath}

\usepackage[T1]{fontenc}
\usepackage{subfig}
\usepackage[export]{adjustbox}
\usepackage{multirow}

\makeatletter \renewcommand\@biblabel[1]{$^{#1}$} \makeatother
\newcommand{\cen}[1]{\begin{center} #1 \end{center}}

\usepackage[mathlines,edtable]{lineno}

\usepackage{hyperref}
\hypersetup{ colorlinks,
    citecolor=blue,
    filecolor=blue,
    linkcolor=blue,
    urlcolor=blue
}

\usepackage{xcolor}

\definecolor{gray}{rgb}{0.6,0.6,0.6}
\definecolor{red}{rgb}{0.85,0,0}
\definecolor{green}{rgb}{0,0.85,0}
\definecolor{blue}{rgb}{0,0,0.85}
\definecolor{beige}{rgb}{0.92,0.87,0.78}
\usepackage[all]{hypcap}    

\begin{document}

\def\bea{\begin{align}}
\def\eea{\end{align}}
\def\beq{\begin{equation}}
\def\eeq{\end{equation}}

\def\ea{\textit{et al.}} 
\def\hf{\mathrm{HF}}
\def\hfcy{\langle\mathrm{HF}_{\mathrm{cyclic}}\rangle}
\def\hfcypot{\mathrm{HF}_{\mathrm{pot.}}}
\def\hfch{\mathrm{HF}_{\mathrm{chronic}}}
\def\sfp{\mathrm{SF}}
\def\sfo{\mathrm{SF}^{(1)}_{\mathrm{oxic}}}
\def\sfch{\mathrm{SF}^{(1)}_{\mathrm{chronic}}}
\def\sfcy{\mathrm{SF}^{(1)}_{\mathrm{cyclic}}}
\def\bed{\mathrm{BED}}
\def\bedo{\mathrm{BED}_{\mathrm{oxic}}}
\def\bedch{\mathrm{BED}_{\mathrm{chronic}}}
\def\bedcy{\mathrm{BED}_{\mathrm{cyclic}}}
\def\bedone{\mathrm{BED}(1)}
\def\bedN{\mathrm{BED}(N)}
\def\bedoneo{\mathrm{BED}_{\mathrm{oxic}}(1)}
\def\bedNo{\mathrm{BED}_{\mathrm{oxic}}(N)}
\def\bedNh{\mathrm{BED}_{\mathrm{hy}}(N)}
\def\pc{p_{\mathrm{c}}}
\def\pcmax{p_{\mathrm{c,max}}}
\def\pcmin{p_{\mathrm{c,min}}}
\def\pcm{\bar{p_{\mathrm{c}}}}
\def\br{\mathbf{r}}
\def\nc{n_{\mathrm{c}}}
\def\ld{l_D}
\def\diso{d_{\mathrm{iso-SF}}}
\def\Diso{D_{\mathrm{iso-SF}}}
\def\bedeff{\mathrm{BED}_{\mathrm{iso-SF}}}

\cen{\sf {\Large {\bfseries A simple mathematical model of cyclic hypoxia and its impact on hypofractionated radiotherapy} \\  
\vspace*{10mm}
Edward Taylor} \\
Princess Margaret Cancer Centre, University Health Network, Toronto,  Canada\\
Department of Radiation Oncology, University of Toronto, Toronto, Canada
\vspace{5mm}\\
January 10, 2023\\
}

\pagenumbering{roman}
\setcounter{page}{1}
\pagestyle{plain}
edward.taylor@rmp.uhn.ca \\

\begin{abstract}
\noindent {\bf Purpose:} There is evidence that the population of cells that experience fluctuating oxygen levels (``acute'', or, ``cyclic'' hypoxia) are more radioresistant than chronically hypoxic ones and hence, this population may determine radiotherapy (RT) response, in particular for hypofractionated RT, where reoxygenation may not be as prominent. A considerable effort has been devoted to examining the impact of hypoxia on hypofractionated RT; however, much less attention has been paid to cyclic hypoxia specifically and the role its kinetics may play in determining the efficacy of these treatments. Here, a simple mathematical model of cyclic hypoxia and fractionation effects was worked out to quantify this.\\
{\bf Methods:} Cancer clonogen survival fraction was estimated using the linear quadratic model, modified to account for oxygen enhancement effects.  An analytic approximation for oxygen transport away from a random network of capillaries with fluctuating oxygen levels was used to model inter-fraction tissue oxygen kinetics. The resulting survival fraction formula was used to derive an expression for the iso-survival biologically effective dose, $\bedeff$. These were computed for some common extra-cranial hypofractionated radiotherapy regimens. \\
{\bf Results:} Using relevant literature parameter values, inter-fraction fluctuations in oxygenation were found to result in an added 1-2 logs of clonogen survival fraction in going from five fractions to one for the same nominal biologically effective dose (i.e., excluding the effects of oxygen levels on radiosensitivity). $\bedeff$'s for most ultra-hypofractionated (five or fewer fractions) regimens in a given tumour site are similar in magnitude, suggesting iso-efficacy for common fractionation schedules. \\
{\bf Conclusions:} Although significant, the loss of cell-killing with increasing hypofractionation is not nearly as large as previous estimates based on the assumption of complete reoxygenation between fractions. Most ultra-hypofractionated regimens currently in place offer sufficiently high doses to counter this loss of cell killing, although care should be taken in implementing single-fraction regimens. 
 \\
 \end{abstract}

\newpage     

\setlength{\baselineskip}{0.7cm}      

\pagenumbering{arabic}
\setcounter{page}{1}
\pagestyle{fancy}

\section{Introduction}
\label{introsec}
 It is well-known that cells at low oxygen tensions (``hypoxia'') are more radioresistant than those at higher oxygen levels, with as much as three times more dose required to achieve the same level of cell killing~\cite{Alper56,Wouters97} and hence, the presence of hypoxia in a tumour likely plays a key role in determining the response to radiation therapy (RT)~\cite{Hill15}. Several experiments have demonstrated, either indirectly~\cite{Brown79,Yamaura79} or directly~\cite{Chan08}, that this reduction in sensitivity is transient, however: cells that are chronically ($\gtrsim$ days) hypoxic exhibit a more modest dose-enhancing factor of $\sim 1.4$ as compared to the factor of 2.5-3 experienced by cells that are transiently hypoxic, likely the result of impaired ability to repair radiation-induced DNA damage~\cite{Chan08}. There is substantial pre-clinical evidence showing that most tumours harbor a radiobiologically significant population of such transiently hypoxic cells, an effect often referred to as acute or cyclic hypoxia~\cite{Bader20}. 
 
Assuming that this population exists in clinical populations (and there is only indirect evidence of this~\cite{Bader20}), we thus hypothesize that it is the maximally radio-resistant cyclically hypoxic component and its kinetics through treatment that determine RT response. Historically, the large OER of the hypoxic compartment was assumed to be mitigated during conventionally fractionated radiotherapy (typically, 20 or more fractions) by ``reoxygenation'', the process by which cells that are hypoxic early in RT gradually become oxic at subsequent fractions, rendering them more radiosensitive~\cite{Kallman86}. Technological advances over the past two decades have led to the adoption of hypofractionated RT regimens such as stereotactic body radiotherapy (SBRT), lasting five or fewer fractions. While some reoxygenation may occur during these abbreviated regimens~\cite{Taylor20}, it is not clear that it is sufficient to mitigate the impact of hypoxia and cyclic hypoxia may play a greater role.

In this paper, we develop a probabilistic model of cyclic hypoxia to quantify its impact on hypofractionated RT. A number of authors have examined the effect of hypoxia on hypofractionated RT; see e.g. Refs.~\cite{Brown10,Ruggieri10, Carlson11,Lindblom14,Jeong17,Guerrero17,Elamir21}. Comparatively little attention seems to have been paid to modelling the specific impact of cyclic hypoxia, however. Popple and colleagues~\cite{Popple02}, while not explicitly studying hypofractionated RT, quantified the influence of cyclic and chronic hypoxia by modelling the former as a completely stochastic population; i.e., by assuming that cells randomly shuttle between oxic and cyclic hypoxia compartments at each fraction. Ruggieri $\ea$ studied the affect of cyclic hypoxia on hypofractionated RT by assuming that cells in the cyclic hypoxia compartment oscillated between oxic and hypoxic states periodically in time with randomly assigned phases~\cite{Ruggieri10}. Toma-Da\c{s}u and colleagues calculated cancer clonogen survival by simulating oxygen diffusion and metabolism from simulated capillary networks; acute hypoxia was simulated by randomly closing a fraction of simulated capillaries between each fraction~\cite{Toma-Dasu09,Lindblom14}. 

Although differing in some details, the approach taken by us here is conceptually similar to that of these last two references~\cite{Toma-Dasu09,Lindblom14}. A key feature of these works is the spatial aspect of their modelling, via calculations of oxygen metabolism and diffusion for simulated vascular architectures. This is important because, in contrast to purely stochastic approaches, cells that are in well-perfused regions are unlikely to oscillate between extremes of hypoxia and oxia between fractions while cells that are far from blood vessels are likewise unlikely to become oxic. The probability that a cell is oxic or hypoxic at a given fraction is thus not independent of the oxygenation status at previous fractions, since both are functions of the local vascular architecture. Mathematically, this lack of complete stochasticity lessens the impact of decreasing fractionation: for completely stochastic oxygenation dynamics, the probability that a given cell is hypoxic--and hence, maximally radioresistant--over the entire treatment diminishes rapidly with increasing fraction number $N$ as $p^N$, where $p$ is the probability that the cell is hypoxic at a single fraction. (This applies not only to cyclic hypoxia but also assumptions of complete ``stochastic'' reoxygenation of a chronically hypoxic compartment between fractions~\cite{Carlson11}). In contrast, modelling the expected non-stochasticity of tissue oxygen tension, the probability that a cell is hypoxic for the duration of RT decays more slowly with $N$. 

The non-stochastic nature of hypoxia dynamics may be of relevance for understanding the apparent success of \emph{ultra}-hypofractionated regimes ($N\leq 5$): if the spatial-temporal kinetics of hypoxia are completely stochastic, one would expect a substantial decrease in tumour control in going e.g., from five fractions to a single one, unless this reduction were accompanied by a substantial increase in dose. Although ultra-hypofractionation has largely been accompanied by increases in doses (as briefly reviewed in Section~\ref{Discussionsec}), these are not sufficient to overcome the loss of cell killing implied by completely stochastic oxygen kinetics. By using a simple approximation to the full oxygen metabolism-diffusion problem~\cite{Toma-Dasu09,Lindblom14} in the present work, we derive analytic expressions for the clonogen survival fractions and iso-survival fraction doses under cyclically hypoxic conditions that we hope will facilitate analysis of fractionation schedules and give insight into the key parameters that control the magnitude of cyclic hypoxia's impact.

We begin by introducing our model of cyclic hypoxia in Section~\ref{cyclicmodelsec} Although it assumes stochastic temporal dynamics in capillary oxygen levels, by incorporating the spatial dependency of hypoxia with respect to the capillary architecture in an approximate way, this model captures the non-stochasticity of tissue oxygen levels described above. In Section~\ref{RTsec}, we combine this model with the linear quadratic model to obtain an analytic expression for clonogen survival as a function of the chosen fractionation regimen and the amplitude of cyclic hypoxia. We close in Section~\ref{Discussionsec} by discussing the implications of our results for some modern hypofractionated SBRT regimens currently in use for lung, liver, pancreas, and prostate cancers, and propose a method to combat cyclic hypoxia.

\section{Cyclic hypoxia model}
\label{cyclicmodelsec}

Acute or cyclic hypoxia arises from transient fluctuations in red blood cell flux~\cite{Chaplin87,Dewhirst96,Kimura96,Lanzen06}. The impact of a reduced blood cell flux through capillaries is to reduce the oxygen tension $\pc$ inside the capillaries~\cite{Kimura96}, the value determined by the interplay between the number of red blood cells entering a region per unit time (flux) and the rate of oxygen metabolism in the tissue there. Fluctuations in $\pc$ yield fluctuations in tissue oxygen levels, yielding cyclic hypoxia. Experiments have shown that cyclic hypoxia is spatially correlated over domains approximately 200-300 $\mu$m in extent~\cite{Lanzen06}, implicating an ``upstream''~\cite{Kimura96} effect such as thermoregulation-induced vasomotion~\cite{Bader20} that simultaneously impacts groups of microvessels. 

A key quantity entering this work is the hypoxic fraction, denoted by $\hf_{\Lambda}$, describing the fraction of tissue for which the local oxygen tension $p(\br)$ at a point $\br$ is less than a radiobiologically relevant  threshold oxygen tension $\Lambda$. Given the spatial extent of cyclic hypoxia correlations, for our purposes, it will be useful to define $\hf_{\Lambda}$ as the fraction of tissue in a pseudo-voxel several hundred $\mu$m's in size. Note that this differs from standard nomenclature, wherein hypoxic fraction refers to the fraction of tissue within an entire tumour for which $p(\br)<\Lambda$. Amongst other factors, the oxygen level $p(\br)$ in tissue is a function of that in capillaries, $\pc$. By defining this ``microscopic'' hypoxic fraction, we can associate a voxel-specific value $\pc(t)$ of the time-varying  capillary oxygen tension to all tissue in that pseudo-voxel and hence, express the hypoxic fraction there as a function of this $\pc(t)$: $\hf_{\Lambda} = \hf_{\Lambda}(\pc)$. Since the oxygen enhancement ratio rises rapidly below 5 mmHg~\cite{Carlson06}, an appropriate choice of cutoff $\Lambda$ should be on the order of several mmHg's.

Now, suppose that $\pc$ cycles (possibly irregularly) between minimum and maximum values of capillary oxygen tension, denoted by $\pcmin$ and $\pcmax$, respectively. The fraction $\hfch$ of tissue that is \emph{chronically} hypoxic refers to the fraction of tissue in a pseudo-voxel that is always hypoxic ($p<\Lambda$):
\beq \hfch \equiv \hf_{\Lambda}(\pcmax).\label{hfchronicdef} \eeq
Correspondingly, the \emph{potential} cyclic hypoxic fraction 
\beq \hfcypot \equiv \hf_{\Lambda}(\pcmin)-\hf_{\Lambda}(\pcmax)\label{hfcycpotdef}\eeq
is defined as the fraction of non-chronically hypoxic tissue that, given enough time, is guaranteed to cycle between oxic ($p>\Lambda)$ and hypoxic ($p<\Lambda)$ states. This quantity in general differs from the \emph{mean} cyclic hypoxic fraction $\hfcy$, which is the expectation value of the fraction of non-chronically hypoxic tissue for which $p(\br)< \Lambda$ \emph{at any point in time}. We assume that the $\pc$ values in each pseudo-voxel and at each time-point (fraction) fluctuate independently and hence, $\pc$ can be modelled as being sampled from a distribution $f(\pc)$, bounded between $\pcmin$ and $\pcmax$. The mean cyclic hypoxic fraction is thus given by
\beq \hfcy = \int_{\pcmin}^{\pcmax}  f(\pc) \hf_{\Lambda}(\pc)\; d\pc -  \hf_{\Lambda}(\pcmax).\label{hfcydef} \eeq
Owing to the differing time scales involved, the potential cyclic hypoxic fraction will play a role in our (brief) discussion of hypoxia binding agents (e.g., pimonidazole and those used in positron-emission tomography imaging of hypoxia), while the mean cyclic hypoxic fraction arises in our model of radiotherapy response.

\begin{figure}
\centering
{{\includegraphics[width=6.24 cm,valign=t]{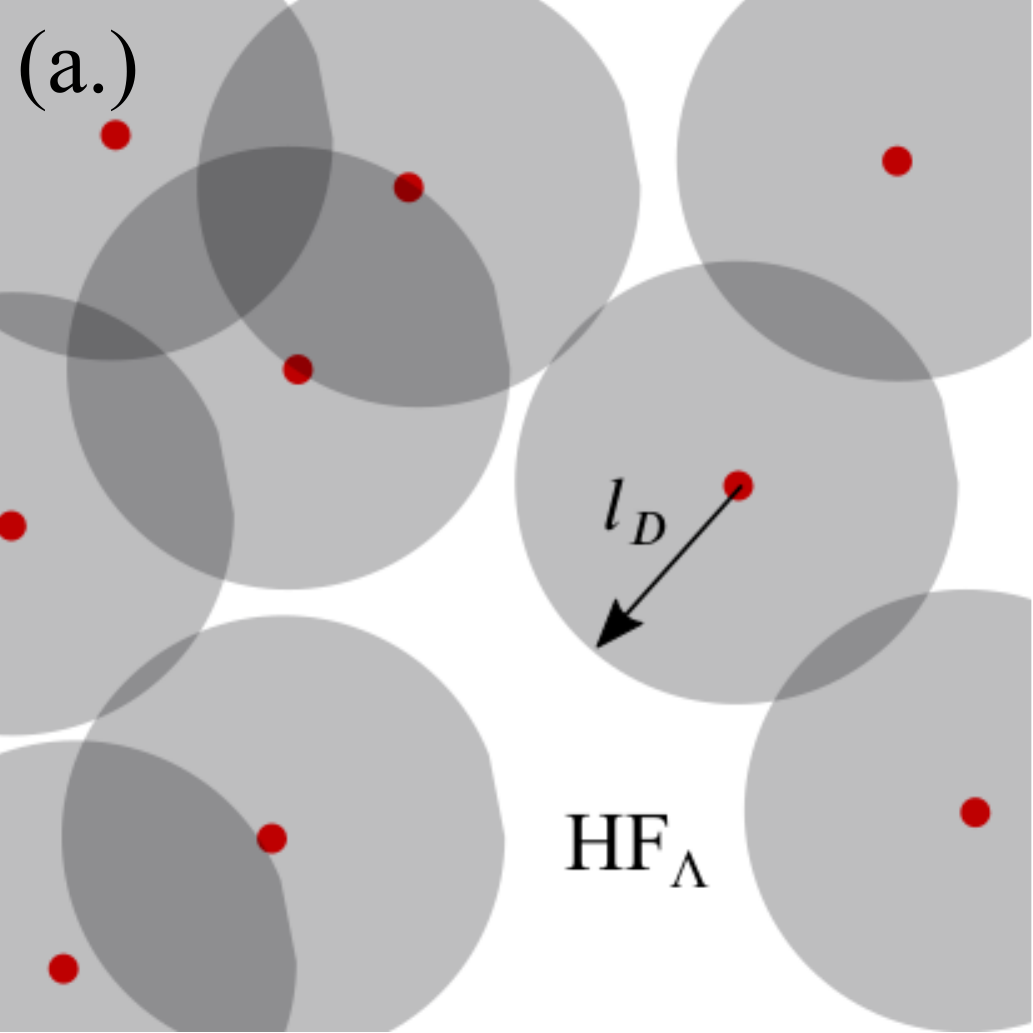}}}
\qquad
{{\includegraphics[width=7.5 cm, valign=t]{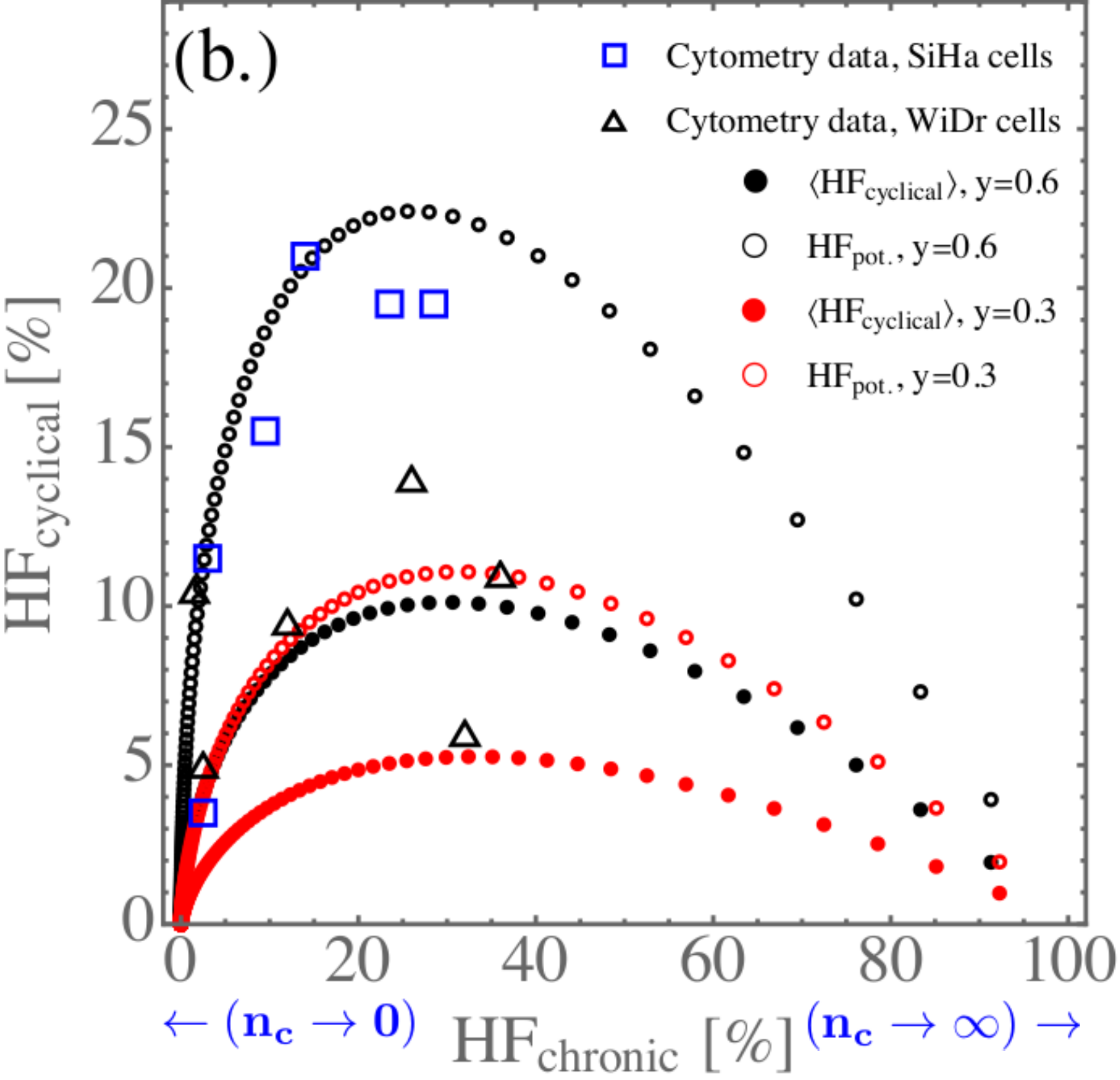}}}
\captionv{12}{}{(a.) Graphical representation of the model shown in Eq.~(\ref{Poissonapprox}). Oxygen diffuses a characteristic distance $\ld$ from capillaries, leaving a fraction $\hf_{\Lambda}$ of tissue hypoxic (unshaded region). For randomly distributed capillaries, this fraction is given by Poisson statistics: $\exp(-\pi n_c \ld^2)$. (b.) Potential cyclical and mean cyclical versus chronic hypoxia, obtained from Eqs.~(\ref{hfchmodel}-\ref{hfcymodel}) for two representative values of the oxygen fluctuation parameter $y\equiv (\Delta p_c/\bar{p}_c) = 0.3$ and 0.6. Also shown is flow cytometry data for two cell lines, extracted from Figs. 2 and 5 of Bennewith and Durand~\cite{Bennewith04} (see text for details). The left and right sides of this plot correspond to the hypo- ($n_c\to 0$) and well-perfused ($n_c\to \infty$) limits, respectively.
\label{hypoxicfxsfig}
}
\end{figure}

The above definitions are completely general. As a specific model of the microscopic hypoxic fraction, consider a randomly distributed vasculature, shown schematically in Fig.~\ref{hypoxicfxsfig}(a.). Oxygen diffuses a characteristic distance $ \ld \equiv \sqrt{4D_{\mathrm{O}_2} \pc/M}$ from each capillary before being completely consumed~\cite{Thomlinson55}. Here,  $D_{\mathrm{O}_2}$ is the diffusivity of oxygen and $M$ is the oxygen consumption rate (mmHg/s, for $\pc$ expressed in mmHg). In a ``micro-region'' of tissue in which these quantities are very nearly constant, the fraction of tissue that is at nearly zero oxygen tension is thus approximated by Poisson statistics:
\beq \hf_{\Lambda}(\pc) \sim e^{-\pi n_c \ld^2} \equiv e^{-\gamma n_c},\;\; \gamma \equiv \frac{4\pi D_{\mathrm{O}_2}\pc}{M},\label{Poissonapprox}\eeq
where $n_c$ is the areal capillary density (number of capillaries per unit area in a given tissue plane) in the pseudo-voxel in question. Equation~(\ref{Poissonapprox}) is an approximation to the solution of the oxygen reaction-diffusion equation with randomly distributed capillaries acting as oxygen sources~\cite{Petit09}. It is approximate since it treats the oxygen ``fields'' [the shaded circles in Fig.~\ref{hypoxicfxsfig}(a.)] emanating from each capillary as independent of each other. In dealing with the distance $\ld$ from a capillary required for the oxygen tension to vanish, Eq.~(\ref{Poissonapprox}) also implicitly assumes that $\Lambda =  0$; formally, nonzero $\Lambda$ can be accounted for by correcting the diffusion distance: $\ld \to \ld (1-\Lambda/\pc)$. We will ignore this small correction and continue to treat $\Lambda$ as small ($\lesssim$ 5 mmHg), but nonzero. 

Cyclic and chronic hypoxia are not independent but are connected by the vascular architecture and capillary oxygen tension kinetics: for a given rate of oxygen metabolism $M$, both the chronic and cyclic hypoxic fractions are functions of the capillary density and the amplitude $\Delta\pc \equiv \pcmax-\pcmin$ of capillary oxygen tension fluctuations. Thus, chronic and cyclic hypoxia go hand-in-hand, primarily functions of the underlying tissue capillary density. Choosing for simplicity a uniform, ``top-hat'' function
\beq
f(\pc) = \begin{cases}
1/\Delta \pc & \mathrm{for} \;\; \pcmin \leq \pc \leq \pcmax \\ 
0 & \mathrm{otherwise}
\end{cases}
\label{pcdistribution}
\eeq
to represent the distribution of capillary oxygen fluctuations in Eqs.~(\ref{hfchronicdef}-\ref{hfcydef}) gives
\beq  \hfch = e^{-x}e^{-z/2}, \label{hfchmodel}\eeq
\beq  \hfcypot = 2e^{-x}\sinh(z/2), \label{hfcypotmodel}\eeq
and
\beq  \hfcy  =e^{-x} \frac{\sinh (z/2)}{(z/2)} - \hfch.\label{hfcymodel}\eeq
Here, we have defined the dimensionless parameters 
\beq x \equiv \bar{\gamma}n_c, \;\; y \equiv  \left(\frac{\Delta p_c}{\bar{p}_c}\right), \;\; \mathrm{and} \;\; z\equiv x\cdot y,\label{dimensionlessdef}\eeq
with $\bar{p}_c$ is the mean capillary oxygen tension, and $\bar{\gamma}\equiv 4\pi D_{\mathrm{O}_2}\bar{p}_c/M$ the mean $\gamma$ value. These parameters are the key variables regulating the magnitude of cyclic hypoxia and its impact on radiotherapy. $x$ describes the efficiency with which the vasculature delivers oxygen: when $x\lesssim 1$, the mean distance $l_D$ travelled by oxygen from capillaries is shorter than the mean distance $\sim 1/\sqrt{n_c}$ between capillaries, and oxygen does not reach all tissue, resulting in substantial $\hfch$. In the opposite regime where, $x\gtrsim 1$, tissue is well-perfused and the chronic hypoxic fraction is small. $y$ quantifies the amplitude of capillary oxygen tension fluctuations. In the limit $y\to 0$ that oxygen tension fluctuations vanish, both terms on the right-hand side of Eq.~(\ref{hfcymodel}) reduce to $\exp(-x)$ and hence, $\hfcy$ vanishes. $z$ is a compound parameter describing both the amplitude of capillary oxygen fluctuations as well as the baseline vascular oxygen deliver efficiency and hence, characterizes the magnitude of tissue oxygen fluctuations that result from fluctuations in capillary tension. 

Any reasonable alternative choice (e.g., log-normal) for the distribution function shown in Eq.~(\ref{pcdistribution}) would yield qualitatively similar results to ours here. We chose the top-hat form since, in addition to yielding analytic results for the hypoxic fractions and survival fractions following RT,  it minimizes the number of parameters in our model.

\subsection{Hypoxia model parameter values}
\label{parametervaluessec}

Using the classic value $\ld \sim 150$ $\mu$m derived by Thomlinson and Gray~\cite{Thomlinson55}, we estimate $\bar{\gamma} = 7\times 10^{-2}$ mm$^2$. In arriving at this value for $\ld$, these authors also assumed $\bar{p}_c \sim 40$ mmHg, the value we adopt here as well. Estimates of the capillary density $n_c$ vary widely~\cite{Couvelard05,Weidner93,Schor98b}, with typical values spanning $(5-500)$ mm$^{-2}$. Thus, $x\equiv \bar{\gamma}n_c$ varies from $\sim 0.1$ to $\sim 10$. Note that vessel quantification studies such as these measure the ``microvascular density'' and do not distinguish capillaries from small arterioles and venules. Thus, these values may overestimate $n_c$.   Measurements of vascular and peri-vascular oxygen tension fluctuations reveal maximum amplitudes $\Delta p_c$ on the order of  $(5-25)$ mmHg~\cite{Kimura96, Dewhirst96} and hence,  $y\equiv \Delta p_c/\bar{p}_c$ varies from $\sim  0.1$ to $\sim 0.6$. For some context, using the representative value $y = 0.4$, for $x=0.1$ (poorly-perfused tissue), $\hfch\sim 0.9$ and $\hfcy \sim 10^{-2}$, while for $x=10$ (well-perfused), $\hfch\sim 10^{-5}$ and $\hfcy\sim 10^{-4}$.

In Figure~\ref{hypoxicfxsfig}(b.), we plot $\hfcy$ and $\hfcypot$ as functions of $\hfch$, using $y = 0.6$ and $y = 0.3$ as illustrative values. In the same plot, we show the cyclic and chronic hypoxic fractions estimated from flow cytometry data acquired by Bennewith and Durand~\cite{Bennewith04} from xenografts exposed to two different hypoxia markers (pimonidazole and CCI-103F) for different durations to distinguish these fractions. The data points shown in this plot were extracted from individual bins for each xenograft--human cervical (SiHa) and colon (WiDr) cancers--, representing different intensities for the fluorescent perfusion dye Hoechst 33342. This way, each datapoint shown in Fig.~\ref{hypoxicfxsfig}(b.) represents a different distance bin from nearest capillaries. As long as the time $T$ that these tracers spend in the blood is greater than several periods of cyclic hypoxia fluctuations, markers such as these serve as ``integral'' ones, labelling all cells that were hypoxic at some point during $T$ and hence, the uptake will be proportional to $\hfcypot$, and not $\hfcy$. We make no attempt to fit this data to our model, but present it here to give an idea of the magnitudes of measured transient and chronic hypoxias. However, we note two properties that our model curves shown in this plot possess that any physiologically valid model as well as empirical data must satisfy: in the completely avascular limit, $n_c\to 0$, any fluctuations in capillary oxygen tension hypoxia will not impact tissue oxygen levels, which remain identically zero: $\hfch\to 1$ and $\hfcy\to 0$. In the opposite limit, of well-vascularized tissue, $n_c\to  \infty$ (in practice, for our parameter choices, $n_c\gtrsim 100$ mm$^{-2}$ suffices), fluctuations in $\pc$ likewise have no impact on the completely oxic tissue: $\hfch \to 0$ and $\hfcy\to 0$. Hence, the exact $\hfcy$ versus $\hfch$ relationship must have the general shape shown in Fig.~\ref{hypoxicfxsfig}(b.).

\section{Cyclic hypoxia and clonogen survival fraction during hypofractionated radiotherapy}
\label{RTsec}

Having introduced our model for cyclic hypoxia [Eqs.~(\ref{hfchmodel}-\ref{hfcymodel})], we now turn to the problem of working out its impact on RT response. Extending the binary decomposition of Carlson \ea~\cite{Carlson11}, tumor clonogens are stratified into three compartments: an oxic, cyclically and chronically hypoxic compartments, each with their own OER: OER$_{\mathrm{oxic}} = 1$, OER$_{\mathrm{chronic}}$ = 1.37, and OER$_{\mathrm{cyclical}} =$ 2.5~\cite{Chan08}. The surviving fraction SF$^{(1)}$ of cancer clonogens after a single fraction of radiotherapy is approximated as
\beq \sfp^{(1)} \simeq  \hfcy \cdot \sfcy + \hfch \cdot \sfch+ \left(1-\hfcy - \hfch\right)\cdot \sfo,\label{SF1}\eeq
where the single-fraction survival fractions in each compartment are given by the linear-quadratic model expressions
\beq \sfp^{(1)}_{i} \equiv e^{-\alpha \cdot \mathrm{BED}_i(1)},\;\; i = \mathrm{oxic},\mathrm{cyclic}, \mathrm{chronic}\label{SFcompartment}\eeq
Here, $\alpha$ is the intrinsic radiosensitivity parameter and 
\beq \mathrm{BED}_i(N)  \equiv \frac{d\cdot N}{\mathrm{OER}_i}\left(1+\frac{d}{(\alpha/\beta)\mathrm{OER}_i}\right)\label{BEDdef}\eeq
is the compartment-specific biologically effective dose (BED) for $N$ fractions~\cite{Carlson06}, with $\alpha/\beta$ quantifying the intrinsic sensitivity to fractionation effects and $d$ the per-fraction dose. In reality, the survival fraction is given by a convolution over a continuous spectrum of oxygen tensions~\cite{Sovik06,Elamir21}. The rapid variation of the OER with respect to oxygen tension at low values ($\lesssim$ 5 mmHg)~\cite{Alper56} justifies the compartmentalization in Eq.~(\ref{SF1}); however, this approximation can lead to quantitative errors in SF~\cite{Dasu09}.

Neglecting proliferation during abbreviated hypofractionated treatments and assuming that there is no reoxygenation, for us meaning that $\hfcy$ and $\hfch$ remain constant through treatment, the survival fraction after $N$ fractions is approximated as
\beq \sfp^{(N)} \simeq \sum_{m=0}^{N} P(m, N-m)( \sfcy)^m(\sfo)^{N-m} + \hfch\cdot \left(\sfch\right)^N.\label{SFN}\eeq
Here, $P(m,N-m)$ is the probability that a non-chronically hypoxic cell is hypoxic for $m$ fractions, and oxic for the remaining $N-m$. Because it excludes the chronically hypoxic population [accounted for in the last term in Eq.~(\ref{SFN})], it satisfies the normalization condition $\sum_{m=0}^N P(m,N-m) = 1-\hfch$. For a single fraction ($N=1$), this probability is simply the mean cyclic hypoxic fraction,
\beq P(1,0) = \hfcy \label{Prelation1} \eeq
and hence, Eq.~(\ref{SFN}) reduces to Eq.~(\ref{SF1}) when $N=1$. 

As noted in the Introduction, for high dose-per-fraction hypofractionated RT regimens, clonogen survival is dominated by the cyclically hypoxic compartment. To see this, compare $d=12$ Gy with a standard conventional regimen dose-per-fraction, $d=2$ Gy. Using $\alpha = 0.4$ Gy$^{-1}$ and $\alpha/\beta = 10$ Gy, for the hypofractionated regimen, $\sfo/\sfcy \sim 10^{-4}$ and $\sfch/\sfcy \sim 10^{-2}$. Conversely, for the conventional regimen, $\sfo/\sfcy\sim 0.5$ and $\sfcy\sim 0.7$.  Thus, for hypofractionated regimens, the chronically hypoxic and oxic compartments can be neglected and survival is dominated by cyclically hypoxic clonogens that remain hypoxic for all $N$ fractions:
\beq \sfp^{(N)} \simeq P(N, 0)(\sfcy)^N.\label{SFN2}\eeq
The leading-order correction to this expression arising from the $P(N-1,1)$ contribution remains less than 10\% for $d\gtrsim $ 6 Gy and hence, Eq.~(\ref{SFN2}) well-describes most hypofractionated regimens of five or fewer fractions. For longer fractionation schedules, in addition to considering all terms in Eq.~(\ref{SFN}), one would have to model reoxygenation and proliferation kinetics, which are beyond the scope of the present work.

It thus remains to derive an expression for $P(N,0)$ to estimate the impact of fractionation on a tumour with a cyclically hypoxic population. Consider a pseudo-voxel for which the capillary oxygen tensions through $N$ fractions of radiotherapy are $\pc^{(1)}, \pc^{(2)},\cdots,\pc^{(N)}$. Assuming that the vasculature remains constant over the $N$ fractions, the fraction $p(N,0)$ of non-chronically hypoxic tissue in this pseudo-voxel that remains hypoxic for all $N$ fractions is
\beq p(N,0) = \hf_{\Lambda}(\mathrm{max}[\pc])-\hfch,\label{pn}\eeq
where $\mathrm{max}[\pc]$ is the maximum of the $\pc^{(1)}, \pc^{(2)},\cdots,\pc^{(N)}$ values. Note that this is not in general equal to $\pcmax$, but is bounded from above by it. Although high doses per fraction ($\gtrsim 8$ Gy) can ablate the vasculature~\cite{Fuks05,Demidov18}, substantial ($\gtrsim 20\%$) losses do not occur until after 1-2 weeks~\cite{Demidov18}, justifying our approximation of a constant vasculature over the hypofractionated schedules of interest to us here. 

Generalizing the result shown in Eq.~(\ref{pn})  to average over all possible realizations of sequences of $\pc$ values [sampled from the distribution $f(\pc)$] by defining $P(N,0)\equiv \langle p(N,0)\rangle$, gives 
\begin{align} P(N,0) =  N\int_0^{\infty} d\pc^{(1)} &f(\pc^{(1)})\hf_{\Lambda}(\pc^{(1)})\Bigg\{\int_0^{\infty} d\pc^{(2)} f(\pc^{(2)})\Theta(\pc^{(1)}-\pc^{(2)})\times \nonumber\\& \cdots \times \int_0^{\infty}d\pc^{(N)} f(\pc^{(N)}) \Theta(\pc^{(1)}-\pc^{(N)})\Bigg\} - \hfch.\label{PN}\end{align}
The factor $N$ in front of the integrals arises since we have arbitrarily chosen $\pc^{(1)}$ to be the maximum value of $\pc$, as enforced by the Heaviside theta functions $\Theta$, and there were $N$ choices for the fraction with the maximum $\pc$ value.

Using Eq.~(\ref{pcdistribution}) in Eq.~(\ref{PN}), the resulting integral can be evaluated analytically, with the result
\beq P(N,0) =  \hfcy \cdot F(z, N)-\hfch\cdot\left[1-F(z,N)\right],\label{PN2} \eeq
where
\beq F(z, N) \equiv  \frac{N\left[\Gamma(N) - \Gamma(N,z)\right]}{z^{N-1}(1-e^{-z})}.\eeq
Here, $\Gamma(N)$ and $\Gamma(N,z)$ are the complete and incomplete Euler gamma functions, respectively. It can be shown that  $F(z, 1) = 1\; \forall z$  and hence, Eq.~(\ref{Prelation1}) is recovered. For nonzero $z$, $F(z,N)$ decreases with increasing number $N$ of fractions, reflecting the increase in clonogen cell-killing (decrease in SF) due to inter-fraction oxygen fluctuations. To express this effect in terms of biologically effective doses, we define an iso-survival fraction BED as:
\begin{align} \bedeff (N)\equiv -\frac{1}{\alpha}\ln (\sfp^{(N)}) & =  \frac{d\cdot N}{\mathrm{OER}_{\mathrm{cyclic}}}\left(1+\frac{d}{(\alpha/\beta)\mathrm{OER}_{\mathrm{cyclic}}}\right) + \frac{1}{\alpha}\ln [P^{-1}(N,0)]
\nonumber\\ 
&\equiv \bed_{\mathrm{cyclic}}(N) + \frac{1}{\alpha}\ln [P^{-1}(N,0)].\label{effBED}\end{align}
When $N=1$, $F(z,1) = 1$ and the iso-SF BED for the hypoxic fraction reduces to $\bedcy$. When $N>1$, the second term on the right-hand side is non-zero and positive; it represents the added BED arising from inter-fraction fluctuations in oxygen levels. 


\begin{figure}
\centering
\includegraphics[width=16cm]{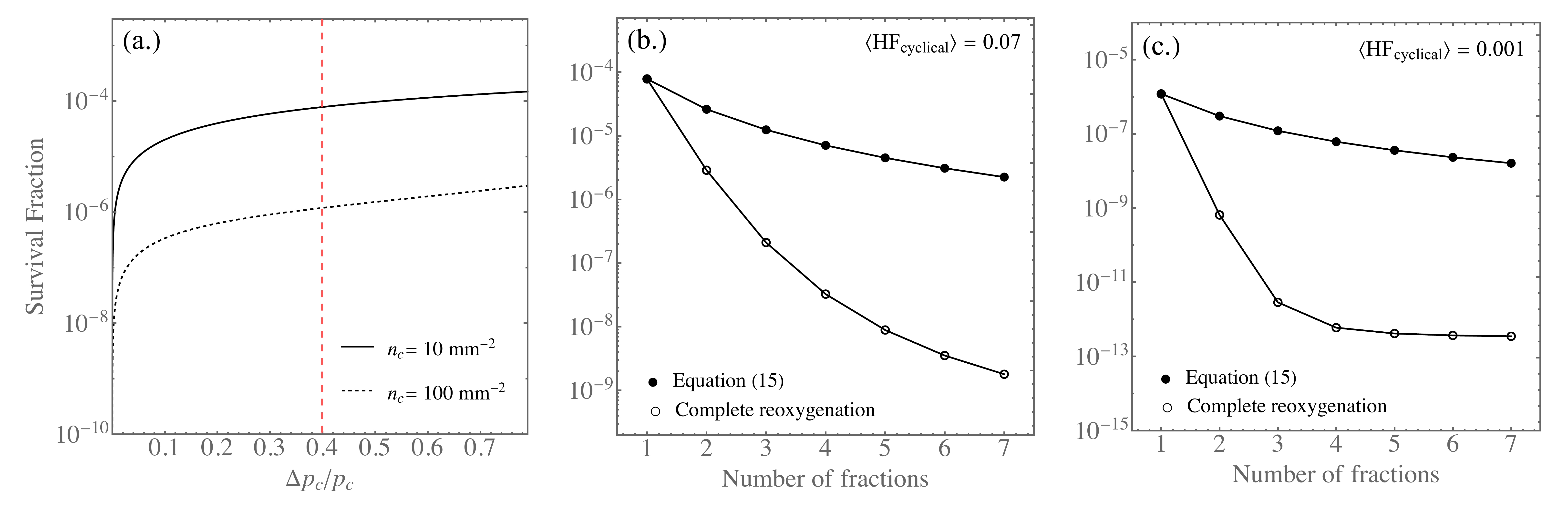}
\captionv{12}{}{Clonogen survival fraction versus (a.) the amplitude $y \equiv (\Delta \pc/\bar{p}_c)$ of capillary oxygen partial pressure fluctuations and (b.), (c.) number of fractions for iso-BED schedules (relative to 60 Gy in 30 fractions, with $\alpha/\beta = 10$).  The dashed-vertical line in (a.) indicates $y=0.4$, the representative value used in the remainder of this manuscript. In (b.) and (c.), SF is shown using this value and $n_c=10$ and $100$ mm$^{-2}$, respectively. Also shown in these plots are the survival fractions assuming complete reoxygenation, given by Eq.~(\ref{reoxy}). For all panels, $\alpha = 0.4$ Gy$^{-1}$ and $\alpha/\beta$ = 10 Gy. 
\label{SFfig}}
\end{figure}

Taken together, Eqs.~(\ref{SFN2}), (\ref{PN2}), and (\ref{effBED}) constitute the primary results of this manuscript. They quantify the impact of cyclic hypoxia on hypofractionated radiotherapy treatments via the estimated fraction of surviving clonogens and the dose needed to counteract the loss of cell-killing due to the reduction in oxygen fluctuations with fewer fractions. The key parameters governing this impact are the fractionation regimen ($d$, $N$) and the compound parameter $z$, which multiplies the vasculature efficiency parameter $x$ and the amplitude $y$ of capillary oxygen fluctuations; see Eq.~(\ref{dimensionlessdef}). In Figure~\ref{SFfig}, using $\alpha = 0.4$ Gy$^{-1}$ and $\alpha/\beta$ = 10 Gy, we plot SF as a function of (a.) the amplitude $y$ of capillary oxygen fluctuations and (b.)-(c.), the number of fractions for nominally iso-BED doses. From Fig.~\ref{SFfig}(a.), it can be seen that, except for very small values of $y$, SF increases slowly with $y$ and we use $y=0.4$ for the remainder of the results in this manuscript, noting the weak dependence of our results on this parameter. Survival fractions in Figs.~\ref{SFfig}(b.) and ~\ref{SFfig}(c.) are plotted as functions of the number of fractions, iso-BED with (the arbitrarily chosen conventional RT schedule 60 Gy in 30 fractions), for two representative values of the capillary density, $n_c = 10$ and 100 mm$^{-2}$. Respectively, these correspond to $x=0.7$ and 7 (poorly- and well-perfused tumours) and cyclic (chronic) hypoxic fractions of 0.07(0.43) and 0.001(0.0002). By ``iso-BED'', we mean in the usual sense of iso-$\bedo$; that is, neglecting OER corrections (by setting OER = 1) in Eq.~(\ref{BEDdef}). For $N=1$, for instance, using $\alpha/\beta=10$, the iso-BED (to 60 Gy in 30) dose is 22.3 Gy. 
The value $\alpha \simeq 0.4$ Gy$^{-1}$ used in these plots is a commonly used value in theoretical radiobiological studies of e.g., head and neck tumours~\cite{Sovik06}, while also being consistent with radiobiological studies of lung cancers~\cite{Jeong17}. $\alpha/\beta =  10$ Gy is a standard value for early-responding (non-prostate) tumour tissues. Also shown in Figure~\ref{SFfig}(b.) and (c.) are the survival fractions when complete, \emph{stochastic} reoxygenation is assumed, corresponding to~\cite{Carlson11}
\beq \sfp^{(N)} = (\sfp^{(1)})^N, \;\;\mathrm{complete}\;\mathrm{reoxygenation}.\label{reoxy}\eeq 

Figure~\ref{SFfig} shows that cyclic hypoxia enhances the efficacy of nominally iso-BED (iso-$\bedo$) radiotherapy with increasing number of fractions, a result of tissue that is initially hypoxic becoming oxic at subsequent fractions due to transient fluctuations. For e.g., $\hfcy$=0.07 [shown in Fig.~\ref{SFfig}(b.)], clonogen cell survival increases by approximately a factor 30  in going from five fractions to a single one.  Although SF increases with decreasing fraction number, the increase is much smaller than arises when complete stochastic reoxygenation occurs between fractions. In the same figure, the assumption of complete reoxygenation leads to a \emph{much} larger increase in SF,  by a factor $\sim 10^5$. 

The expression for the iso-survival fraction biologically effective dose, Eq.~(\ref{effBED}), gives a convenient way of comparing the efficacy of different fractionation schedules. $\bedeff$'s for some common SBRT fractionations schedules are presented in Sec.~\ref{Discussionsec}

\section{Discussion}
\label{Discussionsec}

Driven by advances in technology that have allowed for the reliable delivery of more conformal dose distributions, the use of hypofractionated treatments such as SBRT has grown considerably over the past two decades. Although SBRT has clear efficiency benefits, \emph{prima} \emph{facie}, it is likely suboptimal radiobiologically. Setting aside the differential repair of sublethal DNA damage between early- and late-responding tissues, SBRT challenges two of the ``4 R's of radiotherapy''~\cite{Withers75}: reoxygenation and redistribution through the cell cycle, which enhance radiosensitivity over time. The impact of hypofractionated radiotherapy regimens on tumours with hypoxia has been widely discussed, and the general conclusion has been that tumour control probability will be reduced with fewer fractions because of the reduced extent of reoxygenation~\cite{Brown10,Carlson11}. Why has hypofractionated radiotherapy been so successful then, with no clear evidence of plunging control probabilities even for single-fraction regimens~\cite{Shuryak15}? 

Based on our results, we give two potential reasons for this: first, assuming that reoxygenation does not occur to a substantial degree during hypofractionated RT, cyclic hypoxia was found to result in a more modest decrease in clonogen survival with decreasing fractionation, as compared to models that assume a complete ``stochastic'' reoxygenation; see Fig.~\ref{SFfig}. Second, as we discuss below, in sites such as pancreas and lung, increasing hypofractionation has been accompanied by increasing BED's, largely offsetting the smaller increase in survival predicted here. 

``Complete reoxygenation'' conventionally refers to the phenomenon wherein the hypoxic fraction of \emph{viable} cells (i.e., slated for death but possibly still intact and metabolising) remains approximately constant at each fraction~\cite{Hallbook}. Further assuming that cells are randomly re-assigned to this hypoxic compartment (i.e., stochastic reoxygenation, with probability equal to the hypoxic fraction) at each fraction leads to the result shown in Eq.~(\ref{reoxy}). As was noted in the Introduction, this widely-used approximation likely overestimates the effect of hypofractionation, since it neglects to take into account that the oxygen level kinetics of a given cell depends on its proximity to the nearest capillary and is \emph{not} stochastic. In our model, a cell in close proximity to a blood vessel [i.e., the population of cells not described by Eq.~(\ref{Poissonapprox})] would remain oxic even when the capillary oxygen tension reaches its minimum value (as long as $\pcmin >0$; i.e., $y<1$). Mathematically, this lack of stochasticity is manifested as $P(N,0) \gg \left[P(1,0)\right]^N$ being well-satisfied for $N>1$; that is, the probability that a (non-chronically hypoxic) clonogen will be hypoxic for all $N$ fractions is much greater than $\hfcy^N$. Hence, the decay in cell-killing with decreasing fractionation is much lower than would occur for completely stochastic dynamics.

\begin{table}
\centering
\begin{tabular}{ l | c |  c | c | c }
\hline
\multicolumn{5}{ c }{Biologically effective doses for some common SBRT regimens} \\
\hline
\multirow{2}{*}{
Site }&  $D,N$ & BED$_{\mathrm{oxic}}$ & $\bedeff$  & $\bedeff$ \\ 
& & & ($n_c = 10$ mm$^{-2}$) & ($n_c = 100$ mm$^{-2}$)
\\ \hline
\multirow{4}{*}{Lung} 
& 50, 5~\cite{Takeda09} & 100  & 38 & 50 \\ 
 & 48, 4~\cite{Videtic19} & 106   & 38 &  49 \\
& 34, 1~\cite{Videtic19} & 150 & 39 & 49 \\
& 27*, 1~\cite{Singh19} & 100 &  29 & 40 \\
\hline
\multirow{3}{*}{Pancreas} 
 & 40-50, 5~\cite{Courtney21,Henke18} & 72-100   & 31-38 &  43-50 \\
 & 36, 3~\cite{Mahadevan11} & 79 & 30 & 42 \\
& 25, 1~\cite{Koong04} & 88 & 27 & 37 \\
 \hline
 \multirow{3}{*}{Liver} & 35-40, 5~\cite{Takeda14,Sanuki13} & 60-72  & 29-32 &  43-46 \\ 
 & 40-48, 3~\cite{Bibault13,Andolino11} &93-125 & 34-41& 48-55 \\
 \hline
\multirow{3}{*}{Prostate} 
& 42.7, 7~\cite{Widmark19} &130 &  59 & 105  \\
& 36.25, 5~\cite{Brand19} & 124 &  54 &100  \\
& 19, 1~\cite{Zilli19} & 139 &  45 & 84  \\
\hline
\end{tabular}
\captionv{12}{}{Biologically effective doses for some common SBRT schedules, described by the total dose $D \equiv d\cdot N$ and number of fractions, $N$. Lung cancer and liver schedules refer to early-stage non-small-cell and primary liver cancers, respectively. All BED's are expressed in units of Gy$_{\alpha/\beta}$.  For non-prostate sites, $\alpha = 0.4$ Gy$^{-1}$ and $\alpha/\beta = 10$ Gy were used; for prostate, we used $\alpha = 0.15$ Gy$^{-1}$ and $\alpha/\beta = 3$ Gy~\cite{Nath09}. Iso-SF BED's [Eq.~(\ref{effBED})] are shown for two representative capillary densities and were calculated using $y=0.4$. *Estimated dose accounting for tissue heterogeneity corrections.
\label{BEDstable}
}

\end{table}

Even though our model predicts a much smaller reduction in SF, a 1-2 log increase in SF [e.g., Fig.~\ref{SFfig}(b.)] in going from $\sim$five to one fraction would likely have observable clinical effects except for the fact that, for the most part, hypofractionation has been accompanied by increases in doses and are not ``iso-BED'' (iso-$\bedo$). This is apparent in Table~\ref{BEDstable}, where we list the biologically effective doses--$\bedo$ and $\bedeff$ for two representative values of the capillary density--for some common SBRT regimens. As an example, for early-stage non-small cell lung cancers, in going from 50 Gy in five fractions~\cite{Takeda09} to 34 Gy in  a single fraction (RTOG 0915~\cite{Videtic19}), $\bedo$ increased from 100 Gy$_{10}$ to 150 Gy$_{10}$. At the same time, the iso-SF BED's defined in Eq.~(\ref{effBED}) are nearly identical, reflecting the fact that the 50 Gy$_{10}$ increase in $\bedo$ was just enough to overcome the loss of cell killing due to the diminished effect of inter-fraction oxygen fluctuations.

Although $\bedeff$ depends sensitively on the capillary density $n_c$, the relative $\bedeff$ between different fractions remains largely unchanged. It is also interesting that survival fractions were not overly sensitive to the precise value of the amplitude $y$ of capillary oxygen tension fluctuations [Fig.~\ref{SFfig}(a.)], meaning that almost any (non-zero) degree of cyclic hypoxia would produce comparable $\bedeff$'s. Together, these suggest that $\bedeff$ could be used to compare fractionation schedules, even without knowing $n_c$ or $\Delta p_c$ precisely. 

For most three-to-five fraction regimens in lung, liver, and pancreatic cancers, the estimated iso-SF BED's are nearly the same, providing a possible explanation for why extreme hypofractionation seems to have largely been a success~\cite{Shuryak15,Videtic19}. Primary liver disease is a particularly interesting case, given the apparent absence of a strong dose-response relationship~\cite{Ohri21}. Even though this is undoubtedly due in part to the high rates of local control (LC) in this disease site~\cite{Schaub18}, it is still surprising that even a factor three increase in the nominal BED ($\bedo$) has such little apparent effect on LC.  Accounting for cyclic hypoxia, however, the spread in the iso-SF BED's is greatly reduced, from three to a factor of $\sim$ 1.4.

Nonetheless, our analysis suggests that care should be taken with single-fraction regimens, in particular lung 27 Gy~\cite{Singh19} and prostate 19 Gy~\cite{Zilli19} regimens may be under-dosing tumours when factoring in the impact of cyclic hypoxia. Local control rates and survival at two years have been reported for the lung study and are similar to those in comparable few-fraction regimens~\cite{Singh19}.  It is unlikely, however, that a 1-2 log difference in cell survival would result in a substantial difference in these metrics at two years post-treatment. As an example of this, consider a poorly-perfused lung tumour with $x=0.7$ ($\hfcy = 0.07$). Using Eq.~(\ref{SFN2}), the SF is estimated to be $3\times 10^{-9}$ and $1\times 10^{-7}$ for 48 Gy in four fractions and 27 Gy in  a single fraction, respectively. Now, assuming that some clonogens survived treatment and a tumour doubling time $\tau_{\mathrm{d}}$ of 100 days~\cite{Jeong17}, it would take $\sim$ 6  and $\sim$ 4 years ($= -\tau_{\mathrm{d}} \cdot \ln (\mathrm{SF})/\ln 2$), respectively, for the tumours to regrow to their pre-SBRT sizes. Assessments of treatment response efficacy should thus only be made at least four years after completion of SBRT in sites with high rates of control. Another example is provided by monotherapy prostate brachytherapy, where 19 Gy in a single fraction was found to be inferior to (the iso-$\bedo$ schedule) 27 Gy in two fractions, a difference that was only clearly apparent \emph{after} several years of follow-up~\cite{Morton20}.

Our estimates of clonogen survival and the resulting iso-SF BED are dependent on a number of approximations. The accuracy of the linear quadratic model [Eqs.~(\ref{SFcompartment}) and (\ref{BEDdef})] has been questioned for high doses per fraction~\cite{Kirkpatrick08}. In assessing clinical data, however, others have argued that there is no discernible breakdown of this model~\cite{Brenner08}, in particular when accounting for heterogeneous oxygen distributions that arise in tumours~\cite{Lindblom14b,Jeong17}.  Also, we have not considered the impact of cell-cycle kinetics through radiotherapy, which produce a radio-sensitizing effect with increasing fractionation and which also effect the OER~\cite{Hallbook}. As we have noted throughout this manuscript, the spatial relationship between hypoxia and the vasculature plays a key role in determining the impact of cyclic hypoxia on radiation response since it is responsible for the non-stochasticity of regional levels of hypoxia going though treatment. Our simple expression for this spatial relationship--given by Eq.~(\ref{Poissonapprox})--as well as the compartmentalization used in Eqs.~(\ref{SFN}) and (\ref{SFN2}), are approximations to the continuum distribution of oxygen, obtained by solving the oxygen reaction-diffusion equation for simulated capillary architectures~\cite{Toma-Dasu09,Lindblom14}. We believe, however, that our model captures the essence of cyclic hypoxia and its impact on hypofractionation and that its simplicity will facilitate analyses of SBRT response, in particular for sites where functional imaging is available to assess hypoxia and perfusion.

In addition to solving the full reaction-diffusion equations in their study of cyclic hypoxia and hypofractionated RT, Lindblom and collaborators modelled cyclic hypoxia as a transient, localized occlusion of blood vessels~\cite{Lindblom14}. Although there is significant evidence that this is not the case, and that cyclic hypoxia is caused by transient fluctuations in perfusion~\cite{Dewhirst96,Kimura96,Lanzen06}, this distinction would not qualitatively change the results of our study or the work of Lindblom \textit{et al}.  It would still be the case that cells proximal to capillaries are less likely to find themselves hypoxic at different fractions. Both studies thus find that the impact of hypofractionation is less than would be expected based on the assumption of complete, stochastic, reoxygenation. 

The final major assumption made by us in this work is that substantial reoxygenation (stochastic or otherwise) does not take place during hypofractionated RT. A large body of work has investigated reoxygenation and two types of reoxygenation can be discerned: 1.) transient reoxygenation, occurring up to several hours after radiation, due to transient metabolic or perfusion changes~\cite{Crokart05, Bussink00} and 2.) ``long-term'' ($\gtrsim$ 1 week) reoxygenation kinetics, arising from a reduction in metabolic capacity due to cell death~\cite{Jeong17} and the lysing kinetics of cells killed by RT resulting in blood vessels being brought closer to previously hypoxic cells~\cite{Hallbook}. Both these sources of reoxygenation are unlikely to have a large impact on hypofractionated treatments lasting $\sim 1$ week, with fractions separated by a day or more. Consistent with this, recent pre-clinical measurements of hypoxia using positron-emission tomography in pancreatic tumours revealed only a modest amount of reoxygenation over five fractions (delivered every second day, with hypoxia measurements separated by two weeks)~\cite{Taylor20}. Additionally, Kaleda and colleagues found \emph{increased} levels of hypoxia after (2-4 days post RT) large-dose single-fraction radiotherapy in lung tumours~\cite{Kaleda18}.  

A limitation of the present work is that we have only compared different ultra-hypofractionated radiotherapy regimens with each other, and not with conventionally fractionated regimens. To do so would require modelling reoxygenation~\cite{Jeong17,Guerrero17} and proliferation, as well as using the more general expression for survival fraction given by Eq.~(\ref{SFN}) and hence, computation of $P(m,N-m)$ for $m\neq N$, since these probabilities become non-negligible for smaller per-fraction doses. These straightforward generalizations will be considered in future work.  

In closing, let us ask: are there any actionable discoveries here? Motivated by a desire to understand the apparent success of ultra-hypofractionated radiotherapy regimens, we showed that most of these regimens increase dose sufficiently with decreasing fraction number to overcome the increase in clonogen survival arising from the loss of inter-fraction oxygen fluctuations. Can we offer any recommendations for improvements, then? One possibility concerns ``dose painting'', the delivery of higher doses to more hypoxic regions, typically quantified by positron-emission tomography (PET) using specialized hypoxia-sensitive tracers~\cite{Sovik06}. As with pimonidazole (discussed in Sec.~\ref{parametervaluessec}), these tracers are integral markers of cyclic hypoxia, and tracer uptake in cyclically hypoxic cells will be correspondingly smaller than that in chronically hypoxic ones~\cite{Monnich12}. Because chronically hypoxic cells are also less radioresistant that cyclically hypoxic ones, the survival fraction versus uptake curve (needed for dose painting~\cite{Elamir21}) is relatively ``flat'', restricting the opportunity to meaningfully dose paint based on PET signal. In contrast, our model predicts that the SF versus capillary density ($n_c$) curve will be sharply peaked, reflecting the fact that most radioresistant cyclically hypoxic cells lie in the middle ground between very well-perfused and hypo-perfused; see Fig.~\ref{hypoxicfxsfig}(b). This suggests that patients may benefit from dose painting based on \emph{perfusion} measurements. For our chosen parameter values, SF is maximal for $n_c \sim 10$ mm$^{-2}$ and is decreased by a factor $\sim 10^2$ in going to $n_c \sim 100$ mm$^{-2}$, encompassing a typical range measured in solid tumours~\cite{Weidner93,Schor98b}. Thus, delivery of higher doses to regions with lower $n_c$ (but not $\lesssim 10$ mm$^{-2}$) would decrease clonogen survival. Apart from the inability of hypoxia-PET imaging to resolve cyclic and chronic hypoxia, an advantage of this approach over PET imaging is that qualitative perfusion measurements (e.g., biphasic imaging for liver and pancreatic cancers) are part of routine clinical practice and moreover, even qualitative~\cite{Wang05} metrics have been shown to correlate with microvascular density. Although the precise parameter values (e.g., the rate of oxygen metabolism) required to derive the patient-specific SF versus $n_c$ curve are unknown, the preponderance of perfusion imaging means that it should be possible to derive empirical population-averaged curves by integrating perfusion images with dose distribution maps and outcome data. Future work will explore these directions further. 

\section{Conclusions}
Cells that are transiently hypoxic have been found to be approximately twice as radioresistant as chronically hypoxic cells and between two and three times more radioresistant than oxic cells, possibly due to a diminished capacity to repair DNA damage. Hence, the kinetics of this population through hypofractionated radiotherapy may play a major role in determining treatment response. Using simple analytic approximations for the spatial-temporal dynamics of cyclic hypoxia, we estimated that most ultra-hypofractionated regimens give enough dose to counter losses in clonogen cell-killing with increasing hypofractionation, although some lung and prostate single-fraction regimens may be under-dosing tumours. Dose painting based on perfusion imaging may provide a way to target the cyclically hypoxic population.

\section*{References}
\addcontentsline{toc}{section}{\numberline{}References}
\vspace*{-15mm}



\end{document}